\documentstyle[12pt]{article}
\begin{document}
\title{Self-force approach to synchrotron radiation}
\author{Lior M. Burko\thanks{E-mail: burko@tapir.caltech.edu}\\
Theoretical Astrophysics \\
California Institute of Technology\\
Pasadena, California 91125\\ and \\
Department of Physics\\ Technion---Israel Institute of Technology\\
32000 Haifa, Israel}
\date{\today}
\maketitle

\begin{abstract}
We analyze radiation reaction for 
synchrotron radiation by computing, via a multipole expansion, 
the near field and derive from it the Lorentz four-force, which we  
evaluate on the world-line of the charge. We find that the 
temporal component of the self four-force agrees   
with the radiated power, which one calculates in the radiation zone.  
This is the case for each mode in the multipole decomposition. 
We also find agreement with the
Abraham-Lorentz-Dirac equation. 
\end{abstract}

%PACS numbers: 41.60.Ap\hspace{4.0mm} 04.25-g
%\hspace{4.0mm} 41.60.-m\hspace{4.0mm} 03.50.De

\section{Introduction and overview}
Radiation reaction is a century old problem of physics,   
which has been dealt with via many approaches during the years. (For
reviews see, e.g., \cite{review1,review2,review3}.) Much of the
understanding we currently have on radiation reaction involves
electromagnetic theory \cite{dirac-38,dewitt-brehme},   
but similar problems arise also in other branches of physics, e.g., in
gravitation theory \cite{quinn-wald} (and even in fields such as acoustics 
\cite{templin-99}).

The usual approach for radiation reaction is based on arguments relating
to the balance of quantities, which are constants of motion in the absence
of radiation reaction, specifically the energy and angular momentum. Such
balance arguments involve integration of the flux of an
otherwise conserved quantity over a boundary which normally consists
of a distant sphere. This approach yields the energy and angular momentum
radiated per orbit (i.e, the time-averaged energy and angular momentum
fluxes). Although such methods have been quite useful for the analysis of
many problems involving radiation reaction, they are unsatisfactory for
three main reasons. 
First, they apply only to periodic systems. Systems which involve
radiation reaction are in a strict sense not periodic, because of
the loss of energy through radiation. 
To illustrate the next point, consider a system comprising of a compact
object in motion around a black hole, which is radiating gravitational
radiation. Even if the system is approximated by a quasi-periodic
model, assuming that the radiation reaction effects are small
(specifically, that the time scale for inspiraling is much larger than the
orbital period), eventually the orbit evolution due to radiation reaction
will be fast. Recall now that the usual approach for radiation reaction
yields only the time-averaged energy and angular momentum fluxes.  This
leads us to the second problem with the usual approach: During
the last stages of inspiral the evolution of the orbital motion is fast,
such that the time-averaging methods would be inaccurate.  
The third problem occurs only for systems, which in the absence of
radiation reaction involve non-additive constants of motion. The usual
approach is based on a balance argument for conserved quantities.
Specifically, knowledge of the radiated power at infinity implies that the
system underwent a loss of energy at an equal rate, because of total
energy conservation. However, when non-additive constants of motion are
involved, one can no longer add up the conserved quantity in
the system and that part of the quantity which escaped to infinity, and
then use the conservation law to deduce from the latter on the former.
Such non-additive constants of motion occur for the very interesting case
of motion in the spacetime of a spinning black hole, which is likely to be
observed by future gravitational-wave observatories. 

An alternative approach is to calculate the radiation reaction directly,
at the position of the system, rather than indirectly by balance arguments
and calculation only of the far field in the radiation zone. The first
such method was used by Dirac \cite{dirac-38} for obtaining the
Abraham-Lorentz-Dirac equation for an electric charge in arbitrary motion
in Minkowski spacetime. In that approach one imposes
local conservation laws on a tube surrounding the world-line of the
particle, and integrates the conservation law across the tube, thus
obtaining the equations of motion, including the radiation reaction
effects. However, as is frequently the case with problems involving
radiation reaction, the force which acts on the charge diverges. Dirac 
removed this divergence by using the so-called radiative Green's function,
instead of the retarded Green's function. Specifically, one can write the
retarded Green's function $G^{\rm ret}$ as the sum of two terms, namely  
$$G^{\rm ret}=\frac{1}{2}\left(G^{\rm ret}+G^{\rm adv}\right)
+\frac{1}{2}\left(G^{\rm ret}-G^{\rm adv}\right),$$
where $G^{\rm adv}$ is the advanced Green's function. The first term is
time symmetric. Namely, under the transformation $t\to -t$ the retarded
field transforms to the advanced field and vice versa, such that 
the first term is transformed into itself. Because it is time symmetric,
the first term does not include the radiative part of the field. Instead,
it relates to the non-radiative Coulomb piece of the field, which is the
source for the divergence. Dirac then showed that if one integrates only
the second term, namely the radiative Green's function, one can obtain an
expression for the field of the charge, which is finite on the world line
of the charge. One can then use the usual Lorentz force formula to obtain
the equation of motion which includes also the radiation reaction effects,
i.e., the relativistic Abraham-Lorentz-Dirac equation. Dirac's
prescription for regularization of the self force is quite successful for
arbitrary motion in flat spacetime. However, because it involves the
acausal advanced Green's function, it suffers from an inherent difficulty
when one wishes to apply it for motion in a gravitational field: In curved
spacetime the Green's function has support not only on the light cone, but
also inside it. In particular, the instantaneous forces on the charge
depend on the future history of the charge's motion, which is anti-causal 
\cite{ori-95}. Calculation of the self force which uses only the causal
retarded field should therefore be preferred. 

Recently, Ori proposed a direct method for the calculation of self forces 
based on the retarded Green's function \cite{ori-95,ori-97}.  
The method is based on the
decomposition of the field and the self force into Fourier-harmonic modes.
This approach has two important
advantages: First, when one decomposes the field into modes one can
separate the variables and obtain an ordinary differential equation rather
than a partial one. Second, each mode of the self force turns out to be
finite (even for point-like particles), such that divergences (when occur)
are
met only when one sums over all modes. The treatment of the individual
modes is free from divergences. 
This approach has the advantages of being
both local (such that it does not depend on balance arguments, which are
useless for many interesting cases) and causal (such that it does not
suffer from the acausalities of the radiative Green's function). Of
course, in general the sum over all modes of the self force diverges, and
a regularization
prescription is needed. Such a prescription has been recently proposed by 
Ori \cite{ori-unpublished}. The availability of a regularization
prescription strengthens the motivation for studying 
the self force using this approach. 
For many interesting cases the self force (or some of its components)
is found to be finite without any need of a regularization 
procedure. These cases may serve as an exposition of the direct
calculation of the self force, without the complications which are   
involved with regularization techniques.

As an example for the application of this direct method for the
calculation
of radiation reaction we present in this paper an 
analysis of the radiation-reaction force for (electromagnetic) 
synchrotron radiation. Radiation reaction for this case turns out to
be susceptible of an exact analytical calculation, i.e., one can sum
analytically over all the Fourier-harmonic modes, and obtain a
closed-form expression for the self force.  
In addition, the temporal component of
the self force for synchrotron radiation, whose calculation is our main
objective in this paper, turns out to be finite, without any need of
regularization. 
In fact,  also the angular components of the self force   
turn out to be finite: it is only the radial component which is divergent.
(This divergence can be removed by simple procedures, and the 
renormalized radial self-force is zero.) The separation of the divergent  
component of the self force from the finite components is very useful    
\cite{boyer-72}, as it allows the consideration of important aspects of
the self force without getting involved in complicated regularization
techniques. (In Section \ref{sec7} we shall note in passing on the radial
component of the self force.)

Synchrotron radiation is an invaluable tool in the understanding of
many physical phenomena, and is also very useful for many fields, such as
astronomy, chemistry, biology, medical science, and technology. For recent
reviews see, e.g., \cite{cern-96,margaritondo-88,sokolov-86}. 
The importance of synchrotron radiation stems from its unique properties,
e.g., its spectral distribution in a broad range of wave-lengths, its
relatively high intensity compared with other sources, the high degree of
directionality of the radiation very close to the plane of the charge's 
orbit, and its well-defined polarization. 

The usual approach
to radiation reaction from synchrotron radiation is to compute 
the far field in the radiation zone, then find the asymptotic expression
for 
the Poynting vector and finally calculate the radiated
power by integrating over a large sphere in the radiation
zone. Then one expects the radiated power at infinity to equal the rate
of
loss of energy by the orbiting charge. 
Specifically, one assumes energy balance, and equates the radiated
energy at infinity with the loss of energy by the charge.  
This approach appears in many standard textbooks, e.g., in Refs. 
\cite{jackson,landau}. 
Here we
develop a different approach, which is based on the near field rather
than on the far field.  Specifically, we consider the electromagnetic
field
which the orbiting charge produces, and find the interaction of this field
with the very same charge that created it. Thereby, we construe the
loss of energy by the charge as a self force or a back-reaction
force. Our result here agrees with the well-known result, and also with
the Abraham-Lorentz-Dirac equation \cite{dirac-38}. 
Our method of calculation yields another important
result, which is the multipole decomposition of the radiation. The
radiation has been decomposed in the past into Fourier modes, but no
previous multipole decomposition is known to us. 

We note that our entire discussion here is classical, and that we do not
consider here quantum corrections. There are known cases where
quantum corrections are extremely important, e.g., when there are  
extremely strong magnetic fields, such as those which exist in magnetars.
The importance of 
quantum corrections is discussed in Ref. \cite{erber-83}.
However, for most cases the quantum corrections for synchrotron
radiation are unimportant, such that a purely classical discussion is
useful. We stress that the agreement we find with the known
expressions for synchrotron radiation or with the Abraham-Lorentz-Dirac
equation is found just because we do not consider quantum 
correction. The Abraham-Lorentz-Dirac equation is purely classical, and
does not include any quantum corrections.  

Another point which is worth emphasizing here is that our main result, the
radiated power for synchrotron radiation, can be obtained very easily
using conventional techniques, such as those which appear in Refs. 
\cite{jackson,landau}. Our derivation here is much more complicated from
the technical point of view. However, from the conceptual point of view
it is simple, as it
is, in fact, only a direct calculation of the Lorentz force which uses
only retarded fields. The technical complication is an unavoidable price
which has to be paid in order to use this direct approach. 

We also note that the mathematical formalism which we use is more
sophisticated than the formalism which is usually used for the treatment
of synchrotron radiation. The reason for this is that we consider the
electromagnetic field as derived from a four-vector potential. The
decomposition of the four-vector potential into Fourier-harmonic modes is
possible only in terms of vector spherical harmonics. Vector
spherical harmonics are frequently used in standard texts (e.g., in Ref.
\cite{jackson}). However, 
we preferred to use the conventions of indicial vector
(and tensor) analysis, rather than the notation of \cite{jackson}.  
Because the analysis is done in curved coordinates, we shall frequently
use concepts from Riemannian geometry (or pseudo-Riemannian, in a strict
mathematical sense). The reader is referred to any  text on general
relativity, e.g., Ref. \cite{schutz}. 

The organization of this paper is as follows. 
We begin in Section \ref{sec1} by decomposing the electromagnetic
vector potential into vector spherical harmonics. In the Appendix A we
briefly review the main properties of vector spherical harmonics which are
important to us. More details can be found, e.g., in Ref.
\cite{thorne-80}. 
In Section \ref{sec2} we compute the electromagnetic
field from the vector potential, evaluate the field at the
instantaneous position of the charge, and calculate the temporal component
of the Lorentz four-force. We shall show that it is this component
which is relevant for the radiated power. In Section \ref{sec3} we use
the more conventional approach and decompose the radiated power at
infinity into spherical harmonics. This is done by evaluating the
asymptotic far-field electromagnetic field, building the
energy-momentum tensor from this field, and integrating the
energy flux over a large sphere. This yields the
radiated power at infinity directly, and we show that this far-field
calculation agrees with the self force approach mode by mode. In
Section \ref{sec4} we collect the contributions from the various
multipole moments, and expand the radiated power in a power
series. This series can be summed over all orders, and the result we
obtain for the total radiated power in Section \ref{sec5} coincides
with the well known result, which is obtained, e.g., in Refs.
\cite{jackson,landau} by completely
different procedures. 
We demonstrate agreement with the Abraham-Lorentz-Dirac
equation in Section \ref{sec6}, and discuss our method and results in
Section \ref{sec7}.

\section{The electromagnetic vector potential}\label{sec1}
Let us consider a point-like electric charge $q$ in circular orbit in
Minkowski
spacetime, such that the 
coordinates of the charge, in spherical coordinates, are $r=r_0$, 
$\theta=\pi/2$, $\phi=\phi (t)$, and $t$. The coordinates $r,\theta,\phi,t$ 
are inertial coordinates, e.g., coordinates of an inertial observer at
$r\gg r_0$. The proper time of the charge is $\tau$, such that $t=\gamma\tau$, 
$\gamma$ being the Lorentz factor. We assume the motion of the 
charge is uniform, $\phi (\tau )=\gamma{\bar \omega}\tau$. The angular  
velocity, as measured by a far away inertial observer, is ${\bar \omega}$. 
We take the metric signature to be $(-\; +\; +\; +)$.

\subsection{Temporal component of the vector potential} We separate the
four-vector potential $A^{\mu}$ into a three-vector ${\bf A}$ and a scalar
potential $A^t$. Bold-faced quantities denote three vectors in any basis,
namely either in a covariant, contravariant or in a normalized basis. (See
Appendix A.) We shall use the Lorenz \cite{felsager} gauge
throughout \cite{lorenz}, and use units where
the speed of light $c=1$. 
The scalar potential satisfies the inhomogeneous
wave equation \begin{equation} \Box A^t=-4\pi j^t,
\label{scalar-wave} \end{equation} 
where the temporal component of the four-current density $j^t$ is given by 
$$j^t=\rho
u^t=q\gamma\int_{-\infty}^{\infty}\,d\tau\frac{\delta^{4}[x^{\mu}-
x_{0}^{\mu}(\tau )]}{r^2\sin\theta}.$$ Here, $\rho$ is the charge density,
and $u^t$ is the temporal component of the four-velocity. 
We denote by $\Box$ the d'Alembertian wave operator. 
Decomposition of $j^t$ into Fourier-harmonic modes yields 
\begin{equation}
j^t=q\frac{\delta
(r-r_0)}{r_0^2}\int_{\omega=-\infty}^{\infty} \,d\omega e^{-i\omega
t}\sum_{lm}\delta (\omega-m{\bar \omega})  {\bar P}^{lm}(0)Y^{lm}(\theta
,\phi) .
\end{equation}
We denote ${\bar P}^{lm}(0)\equiv
e^{-im\phi}Y^{lm}(\cos\theta =0 ,\phi)$.  In the Fourier decomposition we
take
the frequencies of the Fourier modes to be both positive and negative. We
next decompose the scalar potential into modes according to  
\begin{equation}
A^{t}=\sum_{lm}\int_{\omega=-\infty}^{\infty} \,d\omega\frac{1}{r}
\psi_{lm,\omega}(r)Y^{lm}(\theta ,\phi)e^{-i\omega t}.  
\end{equation}
Then, the left hand side (lhs) of the wave equation becomes
\begin{equation}
\Box A^{t}=\sum_{lm}\int_{\omega =-\infty}
^{\infty}\,d\omega\left\{\frac{1}{r}\psi ''_{lm,\omega}+ \left[\omega
^2-\frac{l(l+1)}{r^2}\right]\frac{1}{r}\psi_{lm,\omega}
\right\}Y^{lm}e^{-i\omega t}.
\end{equation} 
A prime denotes differentiation with
respect to the radial coordinate $r$. 
Substituting into the wave equation (\ref{scalar-wave})
we find for the modes of the scalar potential:  
\begin{equation}
\psi ''_{lm,\omega}+ \left[\omega
^2-\frac{l(l+1)}{r^2}\right]\psi_{lm,\omega} =-4\pi q\frac{\delta
(r-r_0)}{r_0}\delta (\omega-m{\bar \omega})  {\bar P}^{lm}(0).
\label{eq}
\end{equation}
We choose
the solutions of the corresponding homogeneous equation to be such that
one solution is
regular at the origin (namely, at $r=0$), and the other solution 
describes purely outgoing modes at infinity. These requirements
are equivalent to choosing a 
retarded solution, which is the physically relevant solution. These basic
functions which solve the homogeneous
equation are 
$$\psi_l^0=\sqrt{\frac{|\omega
|}{2\pi}}rj^l(\omega r)$$ and $$\psi_l^{\infty}=\sqrt{\frac{|\omega
|}{2\pi}}rh^l(\omega r),$$ where $j^l,h^l$ are the spherical Bessel and
Hankel functions of the first kinds, respectively. 
Our choice of normalization (which is, of course, not determined from
the homogeneous equations) is the same as in Ref. \cite{thorne-80}. 
The Wronskian determinant \cite{wronskian} of these two solutions is
$W=i\,{\rm sgn}(\omega )/(2\pi )$. 
We show in Appendix B that the solution for Eq. (\ref{eq}) can be written
as a sum of two terms, each proportional to a Heaviside step function.
Specifically, the $lm$ mode of the scalar potential
is given by 
\begin{eqnarray}
A^t_{lm}&=&\frac{f(r_0)\psi^{\infty}_{lm}(r_0)}{W}\frac{1}{r}
\psi^0_{lm}(r)e^{-i\omega t}Y^{lm}(\theta ,\phi )\Theta (r_0-r) \nonumber
\\ &+& \frac{f(r_0)\psi^{0}_{lm}(r_0)}{W}\frac{1}{r}
\psi^{\infty}_{lm}(r)e^{-i\omega t}Y^{lm}(\theta ,\phi )\Theta (r-r_0)
\label{sol} 
\end{eqnarray} 
where $\Theta (x)=1$ for $x>0$ and $\Theta (x)=0$ for $x<0$
is the Heaviside step function, and where $f(r)=-(4\pi
q/r)\delta(\omega-m{\bar\omega}) {\bar P}^{lm}(0)$.

\subsection{Spatial components of the vector potential}
The vector potential ${\bf A}$ satisfies, in the Lorenz 
gauge, an inhomogeneous wave equation
\begin{equation}
\Box {\bf A}=-4\pi{\bf j},
\label{vec-wave}
\end{equation}
where ${\bf j}=\rho{\bf u}$ is the three-current density of the sources
for the field, and where ${\bf u}$ is the three-velocity.  
Decomposing the current density ${\bf j}$ into vector harmonics, we find 
\begin{equation}
j_{\hat i}=q{\bar \omega}\frac{\delta (r-r_0)}{r_0}
\int_{\omega=-\infty}^{\infty}\,d\omega e^{-i\omega t}\sum_{l',lm}\delta 
(\omega-m{\bar \omega}){{\bar P}_{\hat \phi}^{l',lm\; *}}(0)
Y_{\hat i}^{l',lm}(\theta,\phi),
\end{equation}
where ${\bar P}_{\hat \phi}^{l',lm}(0)=e^{-im\phi}
Y_{\hat \phi}^{l',lm}(\cos\theta =0,\phi)$. We denote by a coordinate with
an 
over-hat the coordinate in a normalized basis (see Appendix A), and by a
star we denote complex conjugation. By $Y^{l',lm}_{\hat i}$ we denote a
component in a normalized basis of the pure-orbital vector spherical
harmonics (see Appendix A). We next decompose the vector potential into
modes according to \begin{equation}
A_{\hat i}=\sum_{l',lm}\int_{\omega=-\infty}^{\infty}
\,d\omega\frac{1}{r}
\Psi_{l',lm,\omega}(r)Y_{\hat i}^{l',lm}(\theta ,\phi)e^{-i\omega t}.
\end{equation} 
Then, the decomposition of the lhs of the wave equation into modes yields  
\begin{equation}
\Box A_{\hat i}=\sum_{l',lm}\int_{\omega =-\infty}
^{\infty}\,d\omega\left\{\frac{1}{r}\Psi ''_{l',lm,\omega}+
\left[\omega ^2-\frac{l'(l'+1)}{r^2}\right]\frac{1}{r}\Psi_{l',lm,\omega}
\right\}Y_{\hat i}^{l',lm}e^{-i\omega t}.
\end{equation}
Each mode of the vector potential satisfies the equation 
\begin{equation}
\Psi ''_{l',lm,\omega}+
\left[\omega ^2-\frac{l'(l'+1)}{r^2}\right]\Psi_{l',lm,\omega}
=-4\pi q{\bar \omega}\delta (r-r_0)\delta (\omega-m{\bar \omega})
{\bar P}_{\hat \phi}^{l',lm\; *}(0).
\end{equation} 
The basic solutions of the corresponding homogeneous equation are 
$$\Psi_{l'}^0=\sqrt{\frac{|\omega |}{2\pi}}rj^{l'}(\omega r)$$ and 
$$\Psi_{l'}^{\infty}=\sqrt{\frac{|\omega |}{2\pi}}rh^{l'}(\omega r),$$ 
such that the solution for the $l',lm$ mode of 
the vector potential takes the form 
\begin{eqnarray}
{\bf A}^{l',lm}&=&\frac{F(r_0)\Psi^{\infty}_{l'}(r_0)}{W}\frac{1}{r}
\Psi^0_{l'}(r)e^{-i\omega t}{\bf Y}^{l',lm}(\theta ,\phi )\Theta (r_0-r) 
\nonumber \\ &+&
\frac{F(r_0)\Psi^{0}_{l'}(r_0)}{W}\frac{1}{r}
\Psi^{\infty}_{l'}(r)e^{-i\omega t}{\bf Y}^{l',lm}
(\theta ,\phi )\Theta (r-r_0),
\label{seq}
\end{eqnarray} 
where $F(r)=-4\pi q{\bar \omega}\delta (\omega-m{\bar
\omega})
{\bar P}_{\hat \phi}^{l',lm\; *}(0)$.

We find it convenient to separate the electromagnetic field into electric 
and magnetic modes, which we later find (as indeed should be expected) to 
propagate on their own without mixing. The index $l'$ of the pure-orbital
vector harmonics (do not confuse the prime here with differentiation with
respect to $r$) can take the values $l-1$, $l$, or $l+1$. 
After direct substitution of the explicit  expressions for the various
functions in Eq. (\ref{seq}) and   
summation over all $l'$ modes, 
we find the $lm$ mode of the magnetic part of the vector potential to be 
\begin{equation}
\,{^{M}A}_{\hat t}^{lm}=\,{^{M}A}_{\hat r}^{lm}=0
\end{equation}
\begin{eqnarray}
\,{^{M}A}_{\hat \theta}^{lm}&=&
-a_0^{l,lm}j^l(m\bar \omega r_0)\frac{m}{\sqrt{l(l+1)}}
\left[h^l(m\bar\omega r)e^{-im\bar\omega t}
\frac{Y^{lm}(\theta ,\phi)}{\sin\theta} +c.c\right]
\Theta (r-r_0) \nonumber \\
&-&
a_0^{l,lm} j ^l(m\bar \omega r)\frac{m}{\sqrt{l(l+1)}}
\nonumber \\
&\times &
\left[h^l(m\bar\omega r_0)e^{-im\bar\omega t}
\frac{Y^{lm}(\theta ,\phi)}{\sin\theta} +c.c\right]
\Theta (r_0-r) 
\end{eqnarray}
\begin{eqnarray}
\,{^{M}A}_{\hat \phi}^{lm}&=&
-a_0^{l,lm}j^l(m\bar \omega r_0)\frac{1}{\sqrt{l(l+1)}}
\left[ih^l(m\bar\omega r)e^{-im\bar\omega t}
\frac{\,\partial Y^{lm}(\theta ,\phi)}{\,\partial\theta} +c.c\right]
\Theta (r-r_0) \nonumber \\
&-&
a_0^{l,lm}j^l(m\bar \omega r)\frac{1}{\sqrt{l(l+1)}} \nonumber \\
&\times &
\left[ih^l(m\bar\omega r_0)e^{-im\bar\omega t}
\frac{\,\partial Y^{lm}(\theta ,\phi)}{\,\partial\theta} +c.c\right]
\Theta (r_0-r) 
\end{eqnarray}
and, similarly, for the electric part
\begin{eqnarray}
\,{^{E}A}_{\hat t}^{lm}&=&
\varphi_0 j^l(m\bar\omega r_0)\left[ih^l(m\bar\omega r)e^{-im\bar\omega t}
Y^{lm}(\theta ,\phi)+c.c\right]\Theta (r-r_0) \nonumber \\
&+&
\varphi_0 j^l(m\bar\omega r)\left[ih^l(m\bar\omega r_0)e^{-im\bar\omega t}
Y^{lm}(\theta ,\phi)+c.c\right]\Theta (r_0-r) 
\end{eqnarray}
\begin{eqnarray}
\,{^{E}A}_{\hat r}^{lm}&=&
\left\{ -a_0^{l+1,lm}j^{l+1}(m\bar\omega r_0)\sqrt{\frac{l+1}{2l+1}}
\left[h^{l+1}(m\bar\omega r)e^{-im\bar\omega t }Y^{lm}(\theta ,\phi)
+c.c\right]\right. \nonumber \\
&+&\left. a_0^{l-1,lm}j^{l-1}(m\bar\omega r_0)\sqrt{\frac{l}{2l+1}}
\left[h^{l-1}(m\bar\omega r)e^{-im\bar\omega t }Y^{lm}(\theta ,\phi)
+c.c\right]\right\}\Theta (r-r_0) \nonumber \\
&+&\left\{ -a_0^{l+1,lm}j^{l+1}(m\bar\omega r)\sqrt{\frac{l+1}{2l+1}}
\left[h^{l+1}(m\bar\omega r_0)e^{-im\bar\omega t }Y^{lm}(\theta ,\phi)
+c.c\right]\right. \nonumber \\
&+&\left. a_0^{l-1,lm}j^{l-1}(m\bar\omega r)\sqrt{\frac{l}{2l+1}} \right.
\nonumber \\ &\times& \left.
\left[h^{l-1}(m\bar\omega r_0)e^{-im\bar\omega t }Y^{lm}(\theta ,\phi)
+c.c\right]\right\}\Theta (r_0-r)
\end{eqnarray}
\begin{eqnarray}
\,{^{E}A}_{\hat \phi}^{lm}&=&
\left\{ a_0^{l+1,lm}j^{l+1}(m\bar\omega r_0)\frac{m}{\sqrt{(l+1)(2l+1)}}
\left[ih^{l+1}(m\bar\omega r)e^{-im\bar\omega t }
\frac{Y^{lm}(\theta ,\phi)}{\sin\theta}
+c.c\right]\right. \nonumber \\
&+& \left. a_0^{l-1,lm}j^{l-1}(m\bar\omega r_0)\frac{m}{\sqrt{l(2l+1)}}
\left[ih^{l-1}(m\bar\omega r)e^{-im\bar\omega t }
\frac{Y^{lm}(\theta ,\phi)}{\sin\theta}
+c.c\right]\right\}\Theta (r-r_0) \nonumber \\
&+&\left\{ a_0^{l+1,lm}j^{l+1}(m\bar\omega r)\frac{m}{\sqrt{(l+1)(2l+1)}}
\left[ih^{l+1}(m\bar\omega r_0)e^{-im\bar\omega t }
\frac{Y^{lm}(\theta ,\phi)}{\sin\theta}
+c.c\right]\right. \nonumber \\
&+& \left. a_0^{l-1,lm}j^{l-1}(m\bar\omega r)\frac{m}{\sqrt{l(2l+1)}}
\right. \nonumber \\ 
&\times & \left.
\left[ih^{l-1}(m\bar\omega r_0)e^{-im\bar\omega t }
\frac{Y^{lm}(\theta ,\phi)}{\sin\theta}
+c.c\right]\right\}\Theta (r_0-r)
\end{eqnarray}
\begin{eqnarray}
\,{^{E}A}_{\hat \theta}^{lm}&=&
\left\{ a_0^{l+1,lm}j^{l+1}(m\bar\omega r_0)\frac{1}{\sqrt{(l+1)(2l+1)}}
\left[h^{l+1}(m\bar\omega r)e^{-im\bar\omega t }
\frac{\,\partial Y^{lm}(\theta ,\phi)}{\,\partial\theta}
+c.c\right]\right. \nonumber \\
&+& \left. a_0^{l-1,lm}j^{l-1}(m\bar\omega r_0)\frac{1}{\sqrt{l(2l+1)}}
\left[h^{l-1}(m\bar\omega r)e^{-im\bar\omega t }
\frac{\,\partial Y^{lm}(\theta ,\phi)}{\,\partial\theta}
+c.c\right]\right\}\Theta (r-r_0) \nonumber \\
&+&\left\{ a_0^{l+1,lm}j^{l+1}(m\bar\omega r)\frac{1}{\sqrt{(l+1)(2l+1)}}
\left[h^{l+1}(m\bar\omega r_0)e^{-im\bar\omega t }
\frac{\,\partial Y^{lm}(\theta ,\phi)}{\,\partial\theta}
+c.c\right]\right. \nonumber \\
&+& \left. a_0^{l-1,lm}j^{l-1}(m\bar\omega r)\frac{1}{\sqrt{l(2l+1)}}
\right. \nonumber \\
&\times & \left. 
\left[h^{l-1}(m\bar\omega r_0)e^{-im\bar\omega t }
\frac{\,\partial Y^{lm}(\theta ,\phi)}{\,\partial\theta}
+c.c\right]\right\}\Theta (r_0-r) 
\end{eqnarray}
Here, 
\begin{eqnarray}
a_0^{l+1,lm}&=&\frac{4\pi qm^2\bar\omega^2r_0}{\sqrt{(l+1)(2l+1)}}
{\bar P}^{lm}(0) \\
a_0^{l,lm}&=&-\frac{4\pi qm\bar\omega^2r_0}{\sqrt{l(l+1)}}
\frac{\,\partial{\bar P}^{lm}(0)}{\,\partial\theta} \\
a_0^{l-1,lm}&=&\frac{4\pi qm^2\bar\omega^2r_0}{\sqrt{l(2l+1)}}
{\bar P}^{lm}(0) \\
\varphi_0&=&-4\pi qm\bar\omega{\bar P}^{lm}(0) , 
\end{eqnarray}
where ${\,\partial{\bar P}^{lm}(0)}/{\,\partial\theta}=
{\,\partial Y^{lm}}/{\,\partial\theta}(\theta=\pi /2,\phi=0)$.
The full vector potential is obtained by summing over the electric and the
magnetic contributions, i.e., by  
$A_{\hat \mu}^{lm}=\,{^{E}A}_{\hat \mu}^{lm}+\,{^{M}A}_{\hat \mu}^{lm}$.
By taking the complex conjugate of each component of the vector potential
we summed, in fact, each $m$ with its reciprocal negative counterpart.
Specifically, because complex conjugation is equivalent to the
transformation
$m\to -m$, adding the complex conjugate is equivalent to adding the $-m$
counterpart to each term. Consequently, when we sum below over all modes,
we need only to sum over positive values of $m$. 

We shall need below the following explicit expressions, which are easy to
derive \cite{arfken-85}:
\begin{equation}
{\bar P}^{lm}(0)=(-1)^m\sqrt{\frac{2l+1}{4\pi}}
\frac{\sqrt{(l-m)!(l+m)!}}{2^l\left(\frac{l-m}{2}\right)!\left(\frac{l+m}{2}
\right)!}\cos\left[\frac{\pi}{2}(l-m)\right]
\end{equation}
\begin{equation}
\frac{\,\partial{\bar P}}{\,\partial\theta}^{lm}(0)=
(-1)^{m+1}\sqrt{\frac{2l+1}{4\pi}}
\frac{\sqrt{(l-m)!(l+m+1)!(l+m+1)}}
{2^l\left(\frac{l-m-1}{2}\right)!\left(\frac{l+m+1}{2}
\right)!}\sin\left[\frac{\pi}{2}(l-m)\right]
\end{equation}

\subsection{The $\omega =0$ modes}
We have not considered the possibility that $\omega =0$. This case 
corresponds to the magnetic number $m=0$. In this case the terms  
proportional to $\omega^2$ on the lhs of the wave equations 
(\ref{scalar-wave}) and (\ref{vec-wave}) no longer
appear. 
The independent solutions of the homogeneous equations are now 
$\psi^0_0=r^{l+1}$, $\psi^{\infty}_0=r^{-l}$, 
$\Psi^0_0=r^{l'+1}$, and $\Psi^{\infty}_0=r^{-l'}$. 
The Wronskian determinants are $W=2l+1$ and $W=2l'+1$, respectively. 

We find the vector potential for this case to be
\begin{equation}
A^{{\hat t}\; {lm}}=-\frac{4\pi q{\bar P}^{lm}(0)}{2l+1}
\frac{r_0^{l}}{r^{l+1}}Y^{lm}\Theta (r-r_0)-
\frac{4\pi q{\bar P}^{lm}(0)}{2l+1}
\frac{r^{l}}{r_0^{l+1}}Y^{lm}\Theta (r_0-r)
\end{equation}
\begin{equation}
{\bf A}^{l',lm}=-\frac{4\pi q{\bar P}^{l',lm}_{\hat\phi}(0)}{2l'+1}
\frac{r_0^{l'+1}}{r^{l'+1}}{\bf Y}^{l',lm}\Theta (r_0-r)
-\frac{4\pi q{\bar P}^{l',lm}_{\hat\phi}(0)}{2l'+1}
\frac{r^{l'+1}}{r_0^{l'+1}}{\bf Y}^{l',lm}\Theta (r-r_0).
\end{equation}
It can be readily shown that this four-vector potential does not 
contribute to the temporal component of the self force or to 
the power carried off by the radiation, so we
shall ignore it in the sequel. 

\section{The self force}\label{sec2}
We wish to present a calculation of the radiation-reaction force 
which is based on the 
near field of the charge. We first 
calculate the temporal component of the Lorentz force, namely,
$f^{\hat t}=qF^{\hat t\hat\mu}u_{\hat\mu}$, where $u_{\mu}$ is the 
four-velocity of the charge $q$, and $F_{\alpha\beta}$ is the Maxwell 
field-strength tensor which is produced by the charge itself. 
This expression for $f^{\hat t}$ is then to be 
evaluated on the world-line of the charge. The most convenient way to 
calculate the components of the Maxwell tensor is in covariant components, 
because then we have simply 
$F_{\mu\nu}=A_{\nu ,\mu}-A_{\mu ,\nu}$, and we do not have to worry about 
covariant differentiation. (Although the metric is flat, the spherical 
coordinates we use are curved, and therefore one would in general need to 
use covariant derivatives rather than partial derivatives.) Then, we shall
change bases again to a normalized basis. 
We note that $A_{\phi}=r\sin\theta A_{\hat\phi}$, 
$A_{\theta}=rA_{\hat\theta}$, $A_{r}=A_{\hat r}$, and 
$A_t=A_{\hat t}$. 

After calculating the temporal component of the self four-force we shall
also calculate the radiated power, in order to demonstrate their
compatibility. 
We shall therefore be interested below in the radiated power, specifically 
in 
$-\frac{\,dE}{\,dt}^{lm}$, where $E$ is the energy of the charge. That is, 
$\frac{\,dE}{\,dt}$ is the rate of change of the charge's energy, which is
negative because the charge loses energy by radiating. The radiated power
${\cal P}$ is positive, and by conservation of energy   
${\cal P}=-\frac{\,dE}{\,dt}$.    
Specifically,  
$-\frac{\,dE}{\,dt}=\int T^{t\mu}\,ds_{\mu}=\int T^{tr}\,ds_r$, 
taking the surface of integration to be a large sphere, whose 
normal is radial. Then, $ds_{\hat r}=r^2\,d\Omega$, where 
$\,d\Omega =\sin\theta\,d\theta\,d\phi$. That is, we take for the radiated 
power at infinity 
$-\frac{\,dE}{\,dt}^{lm}=\int T^{\hat{t}\hat{r}\;lm}r^2\,d\Omega$, where 
$T_{\mu\nu}$ is the energy-momentum tensor of the electromagnetic field. 
However, the temporal component of the self four-force, by Newton's law, 
is the rate of change of the temporal component of the four-momentum, 
namely, the time derivative of the energy. In what follows we shall 
calculate both $f^{\hat t}$ on the charge's world-line and the radiated 
power at infinity, and show that they agree.

\subsection{Magnetic modes}
The only non-vanishing components of the four velocity are the temporal
and the azimuthal components. Namely, only $u_t$ and $u_{\phi}$ are not
zero. Because the Maxwell tensor is skew-symmetric, it is clear that only
$F^{t\phi}$ can contribute to the temporal component of the Lorentz force. 
Calculation of this component yields  
\begin{eqnarray}
\,{^{M}F}_{t\phi}^{lm}&=&\,{^{M}A}_{\phi ,t}^{lm}-
\,{^{M}A}_{t,\phi}^{lm} \nonumber \\ 
&=&\,{^{M}A}_{\phi ,t}^{lm} = r\sin\theta \,{^{M}A}_{\hat{\phi} ,t}^{lm}
\nonumber \\ 
&=&-m\bar\omega a_0^{l,lm}r\sin\theta j^l(m\bar\omega r_0)
\left[ih^l(m\bar\omega r)e^{-im\bar\omega t}Y_{\hat\phi}^{l,lm}+c.c\right]
\Theta (r-r_0) \nonumber \\
&-& m\bar\omega a_0^{l,lm}r\sin\theta j^l(m\bar\omega r)
\nonumber \\ &\times &
\left[ih^l(m\bar\omega r_0)e^{-im\bar\omega t}Y_{\hat\phi}^{l,lm}+c.c\right] 
\Theta (r_0-r) 
\end{eqnarray}
When this is evaluated on the charge's world-line, we find that the field 
is continuous at $r_0$. (We find that to be the case, in fact, for the
entire temporal 
component of the self four-force.) 
We thus find that on the charge's world line 
\begin{equation}
\left. \,{^{M}F}_{\hat{t}\hat{\phi}}^{lm}\right|_{\rm charge}
=-2\frac{m\bar\omega}{\sqrt{l(l+1)}}a_0^{l,lm}{j^{l}}^{2}(m\bar\omega r_0)
\frac{\,\partial\bar P}{\,\partial\theta}^{lm}(0).
\end{equation}
Then, the contribution of the magnetic modes to the temporal component of
the self force is given by
\begin{equation}
\left.\,{^{M}f}^{\hat t ,lm}\right|_{\rm charge}
=\left.q\,{^{M}F}^{\hat t\hat{\phi} ,lm}u_{\hat\phi}\right|_{\rm charge}=
2q\gamma r_0\frac{m\bar\omega ^2}{\sqrt{l(l+1)}}
a_0^{l,lm}{j^{l}}^{2}(m\bar\omega r_0)
\frac{\,\partial\bar P}{\,\partial\theta}^{lm}(0),
\end{equation}
where $\left. u_{\hat\phi}\right|_{\rm charge}=r_0\bar\omega\gamma$.
\subsection{Electric modes}
Contributions to the electric modes come from $l'=l+1$ and $l'=l-1$. 
We denote the sum over these two values of $l'$ by $\bar l$. 
\begin{eqnarray}
\,{^{E}A}_{\hat\phi}^{\bar l ,lm}&=&\,{^{E}A}_{\hat\phi}^{l+1,lm}+
\,{^{E}A}_{\hat\phi}^{l-1,lm}\nonumber \\
&=&\left\{\frac{m}{\sqrt{(l+1)(2l+1)}}a_0^{l+1,lm}j^{l+1}(m\bar\omega r_0)
\left[ih^{l+1}(m\bar\omega r)e^{-im\bar\omega t}\frac{Y^{lm}}{\sin\theta}
+c.c\right]\right. \nonumber \\
&+&\left. \frac{m}{\sqrt{l(2l+1)}}a_0^{l-1,lm}j^{l-1}(m\bar\omega r_0)
\left[ih^{l-1}(m\bar\omega r)e^{-im\bar\omega t}\frac{Y^{lm}}{\sin\theta}
+c.c\right]\right\}\Theta (r-r_0)\nonumber \\
&+&\left\{\frac{m}{\sqrt{(l+1)(2l+1)}}a_0^{l+1,lm}j^{l+1}(m\bar\omega r)
\left[ih^{l+1}(m\bar\omega r_0)e^{-im\bar\omega t}\frac{Y^{lm}}{\sin\theta}
+c.c\right]\right. \nonumber \\
&+&\left. \frac{m}{\sqrt{l(2l+1)}}a_0^{l-1,lm}j^{l-1}(m\bar\omega r)
\right. \nonumber \\
&\times & \left. 
\left[ih^{l-1}(m\bar\omega r_0)e^{-im\bar\omega t}\frac{Y^{lm}}{\sin\theta}
+c.c\right]\right\}\Theta (r_0-r)
\end{eqnarray}
Such that the $t\phi$ covariant component of the Maxwell tensor is 
\begin{eqnarray}
\left. \,{^{E}F}_{t\phi}^{lm}\right|_{\rm charge}&=&\left.
\,{^{E}A}_{\phi ,t}^{\bar l ,lm}-\,{^{E}A}_{t, \phi}^{l,lm}
\right|_{\rm charge} \nonumber \\
&=&2a_0^{l+1,lm}\frac{m^2\bar\omega r_0}{\sqrt{(l+1)(2l+1)}}
{j^{l+1}}^2(m\bar\omega r_0){\bar P}^{lm}(0)\nonumber \\ &+&
2a_0^{l-1,lm}\frac{m^2\bar\omega r_0}{\sqrt{l(2l+1)}}
{j^{l-1}}^2(m\bar\omega r_0){\bar P}^{lm}(0)\nonumber \\ &+&
2\varphi_0m{j^{l}}^2(m\bar\omega r_0){\bar P}^{lm}(0)\nonumber \\
&=&\frac{1}{2\pi q\bar\omega}\left\{\left[a_0^{l+1,lm}{j^{l+1}}
(m\bar\omega r_0)\right]^2+
\left[a_0^{l-1,lm}{j^{l-1}}(m\bar\omega r_0)\right]^2\right. \nonumber \\
&-&\left. \left[\varphi_0{j^{l}}
(m\bar\omega r_0)\right]^2\right\}
\end{eqnarray}
and the contribution of the electric modes to the temporal component of
the self force is given by 
\begin{eqnarray}
\left.\,{^{E}f}^{\hat t ,lm}\right|_{\rm charge}&=&
\left.q\,{^{E}F}^{\hat t\hat\phi ,lm}u_{\hat\phi}\right|_{\rm charge}
\nonumber \\
=&-&8\pi\gamma q^2m^4\bar\omega^4r_0^2{{\bar P}^{lm\; 2}}(0)
\frac{l(l+1)}{(2l+1)^2}\left[\frac{1}{l+1}j^{l+1}(m\bar\omega r_0)
\right. \nonumber \\
&-&\left. \frac{1}{l}j^{l-1}(m\bar\omega r_0)\right]^2 
\end{eqnarray}

\subsection{The total force}
Collecting the contributions of the magnetic and the electric modes we find 
that the $lm$ mode of $f^{\hat t}$ is
\begin{equation}
\left.f^{\hat t ,lm}\right|_{\rm charge}
=-8\pi\gamma (qm^2\bar\omega ^2r_0)^2l(l+1){\cal T}^{lm}, 
\label{total_force}
\end{equation}
where 
\begin{equation}
{\cal T}^{lm}=\left[\frac{{\bar P}^{lm}(0)X^{lm}}{2l+1}\right]^2
+\left[\frac{1}{ml(l+1)}\frac{\,\partial {\bar
P}^{lm}}{\,\partial\theta}(0)
j^l(m\bar\omega r_0)\right]^2.
\end{equation}
Here, 
\begin{equation}
X^{lm}=\frac{1}{l}j^{l-1}(m\bar\omega r_0)-\frac{1}{l+1}
j^{l+1}(m\bar\omega r_0).
\end{equation}
In what follows we shall show that this expression coincides with the 
radiated power at infinity, and expand the force into a power series,
which 
will enable us to find out how much radiation there is in each multipole 
moment.

\section{Mode decomposition of the radiated power at infinity}\label{sec3}
We next expand the Maxwell tensor about $r=\infty$, and calculate 
the leading-order term in $r^{-1}$ of the $tr$ component of 
the energy-momentum tensor, which is related to the radiated power. 
In each term of the energy-momentum tensor there is a product of two 
components of the Maxwell tensor. It can be shown that these products 
do not mix fields of different parities or components of different  
multipole moments. Namely, magnetic modes do not mix with electric modes 
(as should indeed be expected), and modes of different values of $l$ or $m$ 
do not mix. By a right arrow 
we denote that the quantity on the left hand side approaches the
expression on the right hand side as $r$ approaches 
infinity. 
We find 
\begin{equation}
\,{^{M}F}_{\hat t\hat\phi}^{lm}\;\;,\;\;\,{^{M}F}_{\hat\phi\hat r}^{lm}
\longrightarrow
(-1)^li^{l+1}a_0^{l,lm}j^l(m\bar\omega r_0)
\frac{1}{\sqrt{l(l+1)}}\frac{1}{r}
\frac{\,\partial Y^{lm}}{\,\partial\theta}e^{im\bar\omega (r-t)}+c.c
\end{equation}
such that
\begin{equation}
\,{^{M}F}_{\hat t\hat\phi}^{lm}\,{^{M}F}^{\hat\phi\hat r,lm}
\longrightarrow 2a_0^{l,lm \;2}\frac{1}{l(l+1)}\frac{1}{r^2}
{j^{l}}^2(m\bar\omega r_0)\frac{\,\partial Y^{lm}}{\,\partial\theta}
\frac{\,\partial Y^{lm\;*}}{\,\partial\theta}+o.t.
\end{equation}
We denote by $o.t$ oscillatory terms, whose contribution to the total
radiated power vanishes after integration over the sphere. 
Similarly,
\begin{equation}
\,{^{M}F}_{\hat t\hat\theta}^{lm}\;\;,\;\;\,{^{M}F}_{\hat\theta\hat r}^{lm}
\longrightarrow
(-i)^{l}a_0^{l,lm}j^l(m\bar\omega r)\frac{m}{\sqrt{l(l+1)}}\frac{1}{r}
\frac{Y^{lm}}{\sin\theta}e^{im\bar\omega (r-t)}+c.c 
\end{equation}
such that
\begin{equation}
\,{^{M}F}_{\hat t\hat\theta}^{lm}\,{^{M}F}^{\hat\theta\hat r,lm}
\longrightarrow 2a_0^{l,lm \;2}\frac{m^2}{l(l+1)}\frac{1}{r^2}
{j^{l}}^2(m\bar\omega r_0)\frac{Y^{lm}Y^{lm\;*}}{\sin^2\theta}
+o.t.
\end{equation}
The total contribution of the magnetic modes to the energy-momentum tensor 
is thus
\begin{equation}
\,{^{M}T}_{\hat t}^{\;\;\;\hat r ,lm}=
\frac{1}{4\pi}\left(
\,{^{M}F}_{\hat t\hat\phi}^{lm}\,{^{M}F}^{\hat\phi\hat r,lm}+
\,{^{M}F}_{\hat t\hat\theta}^{lm}\,{^{M}F}^{\hat\theta\hat r,lm}\right).
\end{equation}
We integrate this component of the energy-momentum tensor over a 
large sphere at infinity. Recalling that 
$$\int\left(\frac{\,\partial Y^{lm}}{\,\partial\theta}
\frac{\,\partial Y^{\tilde{l}\tilde{m}\;*}}{\,\partial\theta}+
\frac{m^2}{\sin^2\theta}Y^{lm}Y^{\tilde{l}\tilde{m}}
\right)\,d\Omega=
l(l+1)\delta_{l\tilde{l}}\delta_{m\tilde{m}},$$
we find
\begin{eqnarray}
{\cal P}^{M\; lm}&=&-\int \,{^{M}T}^{\hat t \hat r ,lm}r^2\,d\Omega =
\frac{1}{2\pi}a_0^{l,lm \;2}{j^{l}}^2(m\bar\omega r_0) \nonumber \\
&=&8\pi q^2\bar\omega^4r_0^2\frac{m^2}{l(l+1)}
{\frac{\,\partial {\bar P}^{lm} (0)}{\,\partial\theta}}^2
{j^{l}}^2(m\bar\omega r_0).
\label{rad-m}
\end{eqnarray}

For the contribution of the electric modes we find
\begin{equation}
\,{^{E}F}_{\hat t\hat\phi}^{lm}\;\;,\;\;\,{^{E}F}_{\hat\phi\hat r}^{lm}
\longrightarrow
-(-i)^l\frac{4\pi qm^3\bar\omega^2}{r}\frac{r_0\bar P^{lm}(0)}{2l+1}
X^{lm}\frac{Y^{lm}}{\sin\theta}e^{im\bar\omega (r-t)}+c.c
\end{equation}
and 
\begin{equation}
\,{^{E}F}_{\hat t\hat\theta}^{lm}\;\;,\;\;\,{^{E}F}_{\hat\theta\hat r}^{lm}
\longrightarrow
-(-i)^{l+1}\frac{4\pi qm^2\bar\omega^2}{r}\frac{r_0\bar P^{lm}(0)}{2l+1}
X^{lm}\frac{\,\partial Y^{lm}}{\,\partial\theta}e^{im\bar\omega (r-t)}+c.c
\end{equation}
such that 
\begin{equation}
\,{^{E}T}_{\hat t}^{\;\;\;\hat r ,lm}=
\frac{1}{4\pi}\left(
\,{^{E}F}_{\hat t\hat\phi}^{lm}\,{^{E}F}^{\hat\phi\hat r,lm}+
\,{^{E}F}_{\hat t\hat\theta}^{lm}\,{^{E}F}^{\hat\theta\hat r,lm}\right),
\end{equation}
and we find
\begin{eqnarray}
{\cal P}^{E\; lm}&=&
-\int \,{^{E}T}^{\hat t \hat r ,lm}r^2\,d\Omega \nonumber \\
&=&8\pi q^2m^4\bar\omega^4r_0^2\frac{l(l+1)}{(2l+1)^2}
{\bar P}^{lm\; 2}(0)\nonumber \\ &\times & 
\left[\frac{1}{l+1}j^{l+1}(m\bar\omega
r_0)-\frac{1}{l}
j^{l-1}(m\bar\omega r_0)\right]^2 .
\label{rad-e}
\end{eqnarray}
The total radiated power for the $lm$ modes is the sum of the magnetic 
and the electric contributions, which coincides with the expression 
above for the temporal component of the self four-force
(\ref{total_force}) up to a
multiplicative factor of $-\gamma$, which arises because the force is
defined by $f_{\mu}=\,dp_{\mu}/\,d\tau$, whereas the radiated power is
equated to $\,dE/\,dt$ . 
The minus sign results from the self force being the rate of change of the
charge's energy which is negative, whereas the radiated power is positive.
We stress that for 
the case of the radiated power we used the far-field expressions for 
the electromagnetic field, namely, the field very far from the charge, 
whereas for the case of the four force we calculated the near-field, 
namely, the interaction of the field which the charge itself creates with 
the very same charge. In both cases we used only retarded fields. 

\section{Power-series expansion of the radiation field}\label{sec4}
In order to gain more insight into the expression we derived for the 
radiated power for each $lm$ multipole moment, and to facilitate the
summation over all modes, 
let us expand our 
expression in a power series. For each multipole mode we find a series
expansion in $\bar\omega^2r_0^2$, such that only even powers
of $\bar\omega r_0$ appear, as should be expected. This expansion is most
easily obtained by using the standard expansions of the spherical Bessel
functions for small values of their arguments. Next, we sum over all modes
$m$ for a given value of $l$.  
The power series
expansion can be obtained for all values of $l$. For example, for the
dipole electric mode ($l=1$) we find for the radiated power:
\begin{equation}
{\cal 
P}^E_1=\frac{2}{3}q^2\bar\omega^4r_0^2\left[1-\frac{2}{5}\bar\omega^2r_0^2
+\frac{43}{700}\bar\omega^4r_0^4+O\left(\bar\omega^6r_0^6\right)\right].
\end{equation} 
This expression reflects the expansion of the radiated power in powers of
the velocity squared. 
Our expressions for the radiated power in each $lm$ mode of
the field [Eqs. (\ref{rad-m}) and (\ref{rad-e})] 
are not restricted to slow motion, and are, in fact, fully
relativistic. We find it convenient to expand in powers of the velocity
(and thus obtain  an expansion where the smallness parameter is the
velocity) because this provides us with a very effective method to sum
over all modes, which is translated into summation over all orders of the
power expansion. When we sum over all orders in the expansion we are no
longer restricted to the slow motion assumption, and we recover the fully
relativistic nature of our analysis. 

Table \ref{table1} summarizes our results for the first three multipole
moments for both electric and magnetic radiative modes, up to
contribution of order $\bar\omega^8r_0^6$. (All higher modes contribute 
only at higher orders in the power-series expansion.) The first column of
the table relates to the radiation mode, where $E_{l}$ means the $l$
multipole moment of electric radiation, and correspondingly $M_{l}$ for
magnetic modes. The second column describes the order of the contribution.
(All modes have a common factor of $(2/3)q^2\bar\omega^4r_0^2$, which we
suppress in this table.) The third column describes the contribution of
the moment in question, and the fourth column says from what value of the
magnetic number $m$ the contribution comes. Recall that we, in fact, sum
over the contributions of $m$ and $-m$ modes, such that the contribution
of the $m$ mode is just half of what Table \ref{table1} shows.     

We find that the dipole magnetic modes do not radiate, and that the
leading magnetic contribution comes from quadrupole radiation. This
can be understood from the symmetry of the problem. Dipole magnetic
radiation would correspond to a continuous current in a circular loop,
say, with time-dependent current, which creates a magnetic moment, the
second time derivative of which does not vanish. 
In our case the current results from the uniform motion of a discrete
point-like charge, for which the second
time derivative of the magnetic moment is zero. 
Therefore, the lowest moment for magnetic modes
is quadrupole and not dipole as is the case for the electric
modes. Because of the vanishing of the dipole magnetic modes, we find
that the magnetic contribution is smaller than the electric
contribution to the radiated power by a factor $\bar\omega^4r_0^4$. 
Whereas the dipole magnetic contribution to the radiated power
vanishes for all orders in $\bar\omega^2r_0^2$, higher multipole
magnetic modes do not. In general, dipole modes should have
contribution of order $\bar\omega^4r_0^2$, and for higher-order
multipoles the leading contribution is smaller by a factor of
$\bar\omega^ {2(l -1)}r_0^{2(l -1)}$. This is the case for both
electric and magnetic modes, but because of symmetry the dipole
magnetic mode does not contribute. Consequently, the leading magnetic
contribution comes from the quadrupole modes. Another point which
should be made is the contribution of the different $m$ moments to a
certain $l$. In the expressions we obtained for the radiated power, we
see that electric modes have a multiplicative factor of ${\bar
P}^{lm\; 2}(0)$, and magnetic modes have a factor of $\frac{\,\partial
{\bar P}^{lm} (0)}{\,\partial\theta}^2$. Consequently, the
contribution of electric modes vanishes whenever $l-m$ is odd, and the
contribution of magnetic modes vanishes whenever $l-m$ is even. We are
now in a position to see why the magnetic modes are so weak: The
dipole mode does not contribute because with $l=1$, and bearing in mind
that $m=0$ modes do not radiate, the only remaining $m$ is $m=1$, but then
$l-m=0$ is even, and the contribution of the dipole magnetic mode
vanishes. Therefore, the leading
magnetic contribution comes from the quadrupole mode. However, the
$m=2$ modes do not contribute to the quadrupole magnetic mode, because
in that case $l-m=0$ is even, and it is just the $m=1$ mode which can
contribute. This results in the vanishing of the term of order
$\bar\omega^6r_0^4$, and the leading magnetic contribution is only of
order $\bar\omega^8r_0^6$.

\section{The total radiated power}\label{sec5}
We next sum all the contributions of the different mutlipole moments (the
first few appear in Table \ref{table1}).
The summation is done in the following way. All the terms in Table
\ref{table1} have a common factor of $(2/3)q^2{\bar \omega}^4r_0^2$. The
orders we shall refer to below are to powers of ${\bar\omega}^2r_0^2$
beyong this common term. At order $1$ only the electric dipole mode
contributes, with coefficient $1$. At order ${\bar\omega}^2r_0^2$ we have
contributions from the electric dipole mode (with coefficient $-2/5$)
and from the electric quadrupole mode (with coefficient $12/5$). The sum
of
these contributions gives a total coefficient $2$. At order 
${\bar\omega}^4r_0^4$ we have contributions from the electric dipole mode 
(with coefficient $43/700$), the electric quadrupole mode 
(with coefficient $-16/7$), the electric octupole mode 
(two terms, one with coefficient $729/140$, and the other with
coefficient $1/2100$), and the magnetic quadrupole mode 
(with coefficient $1/60$). The sum of all these contributions gives a
total coefficient of $3$. The coefficients of higher-order terms can be
found analogously. Collecting all the contributions 
up to order $\bar\omega^4r_0^4$, we  
obtain for the radiated power 
\begin{equation}
{\cal P}=\frac{2}{3}q^2\bar\omega^4r_0^2\left[1+2\bar\omega^2r_0^2+
3\bar\omega^4r_0^4+O\left(\bar\omega^6r_0^6\right)\right] .
\end{equation}
It can be shown (after a tedious calculation, or checking numerically) 
that for the $n$th order in the expansion inside the 
square brackets one finds the 
contribution to be $n\bar\omega^{2n-2}r_0^{2n-2}$, such that the general
expression is
\begin{equation}
{\cal P}=\frac{2}{3}q^2\bar\omega^4r_0^2\sum_{n=1}^{\infty}
n\bar\omega^{2n-2}r_0^{2n-2}.
\end{equation}
This expression equals

\begin{equation}{\cal P}=\frac{2}{3}q^2\bar\omega^4r_0^2\frac{1}
{\left(1-\bar\omega^2r_0^2\right)^2}=\frac{2}{3}
q^2\gamma^4\bar\omega^4r_0^2,
\end{equation}
which is the well-known expression for the radiated power of synchrotron 
radiation.

We showed above that the $lm$ mode of the radiated power 
corresponds to the $lm$ mode of the temporal component of the 
self four-force. In fact, the only difference is an extra factor of 
$-\gamma$, common to all multipole moments. Therefore, summing all 
the components of the self-force we find
\begin{equation}
f^{\hat t}=-\frac{2}{3}q^2\gamma^5\bar\omega^4r_0^2 .
\end{equation}
However, the self force 
\begin{equation}
f^{\hat t}=\frac{\,dp^{\hat t}}{\,d\tau}=
\gamma\frac{\,dp^{\hat t}}{\,dt}=\gamma\frac{\,dE}{\,dt}=
-\gamma {\cal P},
\end{equation}
as we indeed find. Here, $p_{\mu}$ is the four-momentum.

\section{Comparison with the Abraham-Lorentz-Dirac equation}\label{sec6}
We have obtained $f^{\hat t}$ by calculating the Lorentz force acting on
the charge, where the force originates from the electromagnetic field
which the very
same charge creates. On the other hand, the self force on the
charge is given by the Abraham-Lorentz-Dirac (ALD) equation
\cite{rem-ald,dirac-38}, according to which \begin{equation}
f_{\rm ALD}^{\mu}=\frac{2}{3}q^2\left(\frac{\,D^2u^{\mu}}{\,d\tau 
^2}-u^{\mu}
\frac{\,Du_{\alpha}}{\,d\tau}\frac{\,Du^{\alpha}}{\,d\tau}\right) ,
\label{ald}
\end{equation}
where the operator $D$ denotes covariant differentiation. (Recall that
even though spacetime is flat, the coordinates are curved.) We shall next
show that our result coincides with the ALD equation.  
Because the four-velocity is a constant four-vector, its partial
derivative with respect to proper time vanishes. Therefore, the
four-acceleration is given by 
$$a^{\mu}=\frac{\,Du^{\mu}}{\,d\tau}=\frac{\,Du^{\mu}}{\,dx^{\alpha}}u^{\alpha}=
\Gamma^{\mu}_{\alpha\beta}u^{\alpha}u^{\beta},$$
$\Gamma^{\mu}_{\alpha\beta}$ being the Christoffel symbols of the second
kind. It can be readily verified that only the radial
component of the four-acceleration does not vanish. 
Recall that only $u^t$ and $u^{\phi}$ are not zero. In
spherical coordinates, the only non-zero component of the Christoffel
symbols with lower indices $t$ or $\phi$ is 
$\Gamma^{r}_{\phi\phi}=-r\sin^2\theta$. (Note that 
$\Gamma^{\theta}_{\phi\phi}=-\,\sin\theta\,\cos\theta$, but the cosine
vanishes on the charge's world-line, and therefore this component does not
contribute to the four-acceleration.) Bearing in mind that $\sin\theta=1$
on the world-line of the charge, 
we find that $a^r=-r_0\gamma ^2\bar\omega ^2$, such that the
squared four-acceleration is $a_{\mu}a^{\mu}=r_0^2\gamma^4\bar\omega^4$. 
Substitution in the ALD equation (\ref{ald}) 
yields $f_{\rm ALD}^{\hat t}=-(2/3)q^2\gamma ^5\bar\omega^4r_0^2$, namely, 
we find that 
\begin{equation}
f_{\rm ALD}^{\hat t}=f^{\hat t} .
\end{equation} 
Consequently, not only does our method for calculating the self force (by
means of finding the Lorentz force acting on the charge by the field the
very same charge produces) agree with the known expression for
the radiated power at infinity, it also agrees with the
Abraham-Lorentz-Dirac equation (\ref{ald}).  

\section{Discussion}\label{sec7}
We presented a calculation for the temporal component of the self force
for an electric
charge in circular orbit in Minkowski spacetime. The calculation is based
on decomposing the field which the charge creates into Fourier-harmonic 
modes, and finding 
the self force mode by mode.  We 
derive the Lorentz force, and evaluate it on the world line of the charge. 
In the case of the temporal component it is possible to sum over all modes
analytically. It turns out that this sum is finite, and is equal to the
power
carried by the emitted radiation to infinity. We also find agreement with
the Abraham-Lorentz-Dirac equation. 

We note that even though it appears that the agreement between the self  
force we calculated and the radiated power at infinity is instantaneous, 
we in fact expect both quantities to agree in general only after averaging
over a full cycle. We find in this case the agreement to be instantaneous
because the problem is stationary. Consequently, it happens that the
instantaneous value of the force is also the average over a cycle. In
time-dependent problems we do not expect the agreement to be
instantaneous. We still expect both quantities to agree over a full
period, or after integration over all time if the motion is non-periodic.

One might be surprised that the locally-derived self force turned out to 
be finite without any need of regularization. In general one would indeed
expect the self force to diverge. However, it is reasonable to expect the
self force to be the sum of a divergent piece and the finite, physical
piece. The divergent piece is believed to be proportional to the
acceleration. In our case, only the radial component of the acceleration
is not zero. In particular, the temporal component of the acceleration    
vanishes, such that one can expect the divergent piece of the temporal    
component of the self force to vanish. This left us only with the finite  
piece, which is the physically important one.

We briefly remark on the spatial components of the self four-force. From
symmetry, it is expected that the $\theta$ component should vanish. This
is indeed what we find after a detailed calculation, similar to the
calculation of the $t$ component. The $\phi$ component is proportional to
the $t$ component. Specifically, 
\begin{eqnarray}
f^{\hat \phi}&=&qF^{{\hat\phi}{\hat\mu}}u_{\hat\mu}=
qF^{{\hat\phi}{\hat t}}u_{\hat t}\nonumber \\
&=&-qF^{{\hat t}{\hat \phi}}u_{\hat t}=
-\frac{1}{{\bar \omega}r_0}qF^{{\hat t}{\hat \phi}}u_{\hat \phi}
\nonumber \\
&=&-\frac{1}{{\bar \omega}r_0}f^{\hat t}. \nonumber
\end{eqnarray}
This component of the self force is the component which was calculated in 
Ref. \cite{boyer-72}. We note that one can ask the following question:
What external force should be exerted on a {\it charged} particle, in
order to keep it in uniform circular motion? In addition to the mechanical
centripetal force, there is also another force, which balances
the electromagnetic self force. 
The additional external force which should be applied is
just the minus of the $\phi$ component of the self force we calculated
here. (See also Ref. \cite{boyer-72}.)  

We also calculated the radial
component of the self force using a similar approach. In that case,
however, we were unable to sum over all modes analytically. This
summation can still be done numerically. We find, that this sum diverges,
although each mode is finite. The divergence can be removed, and the
renormalized result agrees with the Abraham-Lorentz-Dirac result for the
radial component, which is zero. We used two regularization methods for
the radial component: First, we model the charge to be extended, and then
we consider the limit of the size of the charge approaching zero. In a
sense, this approach is in the spirit of the classical electron models of
Abraham, Lorentz, Poincar\'{e}, and others. The simplest model for a
spatially-extended particle is a dumbbell model, where the
axis of the dumbbell is radial. (We do not argue that this is a realistic
model for a charge. One hopes, nevertheless, that the self force on a
point-like
charge would be independent of the model used in the limit of vanishing
extension. In addition, the self force 
on a general extended object may be obtained by summing on all 
pairs of particles, where each pair is, in fact, a dumbbell.) 
After a simple
mass renormalization (the bare mass is proportional to the acceleration of
the dumbbell's center, and inversely proportional to the
separation distance between the two edges of the dumbbell
\cite{griffiths-owen}), one indeed
obtains the expected result of zero radial force. A second renormalization
method is based on the observation that the $l$ mode of the radial force
on a point-like charge approaches a constant as $l\to\infty$. When one
subtracts this asymptotic constant from each mode, the sum over all modes
converges to the correct result. 
This latter regularization procedure is a direct application of Ori's
prescription \cite{ori-unpublished}, which is expected to be successful
also for more complicated and interesting cases. 

\section*{Acknowledgments}
I have benefited from invaluable discussions with Amos Ori. I thank Kip
Thorne for reading the manuscript and for useful comments. At Caltech this
research was supported by NSF grant AST-9731698 and NASA grant NAG5-6840. 

\section*{Appendix A: Vector spherical harmonics}
The electromagnetic vector potential is a four vector, and therefore it
cannot be decomposed into the usual scalar spherical harmonics, but
instead needs to be decomposed into vector harmonics. However, because the
Minkowski spacetime is stationary (the metric functions are not functions
of time), and thanks to the fact that the metric is diagonal, one can
split the four vector potential into a scalar (the temporal component) and
a three vector (the spatial components). The latter can then be decomposed
into the usual vector spherical harmonics, whereas the former into the
usual scalar spherical harmonics. Vector spherical harmonics are dealt
with extensively in Ref. \cite{thorne-80}. Here we describe just their
properties which are the most relevant ones for our needs. We shall use
the notation and convention of Ref. \cite{thorne-80}. We shall also give
the relation of the vector spherical harmonic we use to those given by
Ref. \cite{jackson}. 

In order to decompose a general three-vector in flat spacetime, let us
project it onto a
tangent space to a sphere. Because this tangent space is two dimensional,
we need two base vectors in order to span it. These base vectors are most
conveniently chosen to be the gradient of the usual scalar harmonics, and
its dual, which in two dimensions is also a vector \cite{sandberg-78}. 
These are the
Regge-Wheeler harmonics \cite{regge-wheeler}, defined by 
$$\psi_{a}\equiv a_l\frac{\partial}{\,\partial x^{a}}Y^{lm}$$ and  
$$\phi_{a}\equiv b_l\epsilon^{b}_{a}\frac{\partial}{\,\partial
x^{b}}Y^{lm}$$
where $a_l$ and $b_l$ are some normalization factors, which we find
explicitly below. 
The tensor $\epsilon^{b}_{a}$ can be represented by the matrix 
$$\epsilon^{b}_{a}=\left(\begin{array}{cc}
0 & -1/\sin\theta \\
\sin\theta & 0 
\end{array} \right),
$$ 
where the coordinates $x^a$ and $x^b$ can take the values 
$\theta$ or $\phi$.
The constants $a_l$ and $b_l$ can be found as follows. We take the
Regge-Wheeler harmonics $ \psi_{a}$ and $\phi_{a}$, and integrate
their
scalar product over the sphere. One finds 
$$\int\psi_{a}^{lm\;
*}\psi_{b}^{\tilde{l}\tilde{m}}g^{ab}\,d\Omega=
a_{l}^*a_ll(l+1)\delta_{l\tilde{l}}\delta_{m\tilde{m}}$$ and 
$$\int\phi_{a}^{lm\;
*}\phi_{b}^{\tilde{l}\tilde{m}}g^{ab}\,d\Omega=
b_{l}^*b_ll(l+1)\delta_{l\tilde{l}}\delta_{m\tilde{m}},$$ 
and normalization yields
$a_l=b_l=\frac{1}{\sqrt{l(l+1)}}$. 
The metric $g_{ab}$ is the metric on the two-sphere, namely 
$\,d\Sigma ^2=g_{ab}\,dx^a\,dx^b=r^2(\,d\theta^2+\sin ^2\theta\,d\phi
^2)$. 
The cross product of $\psi_{a}$ and
$\phi_{a}$ vanishes, namely 
$$\int\psi_{a}^{lm\;
*}\phi_{b}^{\tilde{l}\tilde{m}}g^{ab}\,d\Omega =0 .$$  

Next, because the metric is diagonal, we can 
explicitly list the components of the Regge-Wheeler harmonics in a
normalized basis [namely, in a basis defined by
$V_{\hat\mu}=\sqrt{\left|g^{\mu\mu}\right|}V_{\mu}$ (no summation over
repeated indices) , where $V_{\mu}$ is any four-vector written in a
covariant basis]: 
 
$$\psi_{\hat\theta}=\frac{1}{\sqrt{l(l+1)}}\frac{\,\partial Y^{lm}}
{\,\partial\theta}$$

$$\psi_{\hat\phi}=\frac{1}{\sqrt{l(l+1)}}\frac{im}
{\sin\theta}Y^{lm}$$

$$\phi_{\hat\theta}=-\frac{1}{\sqrt{l(l+1)}}\frac{im} {\sin\theta}Y^{lm}$$

$$\phi_{\hat\phi}=\frac{1}{\sqrt{l(l+1)}}\frac{\,\partial Y^{lm}}
{\,\partial\theta} .$$

We next denote $\psi_{\hat a}$ and $\phi_{\hat a}$ as 
$Y^{E\; lm}_{\hat a}$ and $Y^{B\; lm}_{\hat a}$, respectively, 
where $x^a$ can take
the values $\theta$ or $\phi$. In addition to the tangent space, we also
have the radial direction, which is orthogonal to the tangent space.
However, the projection onto the radial direction would simply be just the
scalar harmonics. Namely, we have $Y^{R\; lm}_{\hat r}=Y^{lm}$. 
$Y^E$, $Y^B$, and $Y^R$ are called collectively the pure-spin vector
spherical harmonics.

Finally, we find it most convenient for our needs to define
the pure-orbital
vector harmonics, because they are eigen-functions of the flat spacetime
Laplacian operator. These pure-orbital vector harmonics are built from the
pure-spin harmonics in the following way:

$$Y^{l+1, lm}=\frac{1}{\sqrt{2l+1}}\left(\sqrt{l}Y^{E\; lm}-
\sqrt{l+1}Y^{R\; lm}\right)$$

$$Y^{l,lm}=-iY^{B\; lm}$$

$$Y^{l-1, lm}=\frac{1}{\sqrt{2l+1}}\left(\sqrt{l+1}Y^{E\; lm}+
\sqrt{l}Y^{R\; lm}\right)$$

Namely, we find that $Y^{l,lm}$ is purely transverse, whereas the other
two harmonics are two independent combinations of transverse and radial
components. Table \ref{table2} gives the explicit expressions for the
components of the
pure-orbital vector spherical harmonics, in a normalized basis.

For completeness, we also give the relation between the pure-spin vector
spherical harmonics and the vector spherical harmonics which are given,
e.g.,  in Ref. \cite{jackson} and involve the operator 
${\bf L}=(1/i){\bf r\times\nabla}$. First, we define 
$${\bf X}^{lm}=\frac{1}{\sqrt{l(l+1)}}{\bf L}Y^{lm}.$$
[This definition is identical to Eq. (16.43) of Ref. \cite{jackson}.] 
Then, 
$${\bf Y}^{B\; lm}=i{\bf X}^{lm},$$  
$${\bf Y}^{E\; lm}=-i{\bf \hat r}{\bf \times}{\bf X}^{lm},$$
and
$${\bf Y}^{R\; lm}={\bf \hat r}Y^{lm},$$
where ${\bf \hat r}$ is a unit vector in the radial direction.
  
It is useful to calculate the divergence of the pure-orbital vector
harmonics. We find 
$${\bf \nabla\cdot Y}^{l',lm}=-(l'-l)^2\left[l'+\frac{1}{2}(l'-l+1)\right]
\sqrt{\frac{l'+(l'-l)(l'-l+1)/2}{2l+1}}\frac{1}{r}Y^{lm},$$
where $l'$ takes the values $l-1$, $l$, or $l+1$. The pure-spin harmonics
satisfy
$${\bf \nabla\cdot Y}^{R\; lm}=\frac{2}{r}Y^{lm}$$
$${\bf \nabla\cdot Y}^{B\; lm}=0$$
$${\bf \nabla\cdot Y}^{E\; lm}=-\frac{\sqrt{l(l+1)}}{r}Y^{lm}.$$
With these expressions for
the divergence of the vector spherical harmonics, it can be readily shown
that the Lorenz gauge condition is
satisfied separately by each multipole moment of the field, by a direct
substitution of the expressions we found for the components of the
electromagnetic vector potential.

\section*{Appendix B: Solution of a differential equation with a delta
function source}

In this Appendix we shall show how one can find the solution to the
inhomogeneous second order ordinary differential equation 
$$g(x)y''(x)+q(x)y'(x)+h(x)y(x)=f(x)\delta(x-x_0), $$
when the two independent solutions $u(x)$ and $v(x)$ 
of the corresponding homogeneous
equation $g(x)y''(x)+q(x)y'(x)+h(x)y(x)=0$ are known. By a prime we denote
in this Appendix differentiation with respect to $x$. 

We first assume that because the source has a delta function form, we
shall have one functional form for the solution for $x>x_0$, and another
form for $x<x_0$. Specifically, let us seek a solution of the form 
$$y(x)=Au(x)\Theta(x-x_0)+Bv(x)\Theta(x_0-x).$$ Here, $\Theta(x)$ is the
Heaviside step function, which satisfies $\Theta(x)=1$ for $x>0$ and 
$\Theta(x)=0$ for $x<0$. The problem then is to find the constant
coefficients $A,B$, such that we shall find the unique solution for
the given inhomogeneous equation. Let us denote $A(x)\equiv
A\Theta(x-x_0)$ and $B(x)\equiv B\Theta(x_0-x)$. Then, the solution is of
the form $y(x)=A(x)u(x)+B(x)v(x)$. Differentiating $A(x)$ and $B(x)$ we
find that $A'(x)=A\delta(x-x_0)$ and $B'(x)=-B\delta(x_0-x)$. 
Consequently, 
$$y'(x)=A(x)u'(x)+A'(x)u(x)+B(x)v'(x)+B'(x)v(x).$$
According to the variation of the parameters method, we now demand that 
$A'(x)u(x)+B'(x)v(x)=0$. Substituting the expression for $A'(x)$ and 
$B'(x)$ we demand that $A\delta(x-x_0)u(x)-B\delta(x_0-x)v(x)=0$, 
or $[Au(x_0)-Bv(x_0)]\delta(x-x_0)=0$. Specifically, we find that 
$B=Au(x_0)/v(x_0)$.

Then, we find that 
$$y'(x)=A(x)u'(x)+B(x)v'(x)=Au'(x)\Theta(x-x_0)+Bv'(x)\Theta(x_0-x)$$
and
$$y''(x)=Au'(x)\delta(x-x_0)+Au''(x)\Theta(x-x_0)
-Bv'(x)\delta(x_0-x)+Bv''(x)\Theta(x_0-x).$$

We next substitute these expressions in the differential equation. We use
now the fact that the functions $u(x)$ and $v(x)$ solve the homogeneous
equation to eliminate all the terms in the equation which are proportional
to the Heaviside step function. We are thus left with 
$$g(x)Au'(x)\delta(x-x_0)-g(x)Bv'(x)\delta(x-x_0)-f(x)\delta(x-x_0)=0.$$
The solution for this is 
$$Au'(x_0)-Bv'(x_0)=f(x_0)/g(x_0).$$ 
Recalling that $B=Au(x_0)/v(x_0)$ we readily  find that 
$$A=\frac{f(x_0)}{g(x_0)} \frac{v(x_0)}{v(x_0)u'(x_0)-u(x_0)v'(x_0)}=
\frac{f(x_0)}{g(x_0)} \frac{v(x_0)}{W(x_0)},$$
where $W$ is the Wronskian of the two functions $v$ and $u$. We thus find
that $A=[f(x_0)/g(x_0)]v(x_0)/W(x_0)$, and 
$B=[f(x_0)/g(x_0)]u(x_0)/W(x_0)$, such that the solution for the
differential equation is given by 
$$y(x)=\frac{f(x_0)}{g(x_0)} \frac{v(x_0)}{W(x_0)}u(x)\Theta(x-x_0)+
\frac{f(x_0)}{g(x_0)} \frac{u(x_0)}{W(x_0)}v(x)\Theta(x_0-x).$$

\newpage
\begin{table}
\begin{tabular}{||c|c|c|c||} \hline
Mode & Order & Contribution & Value of $m$ \\ \hline
   & 1 & 1 & 1 \\ \cline{2-4}
$E_1$   & $\bar\omega^2r_0^2$ & $-2/5$ & 1  \\ \cline{2-4}
   & $\bar\omega^4r_0^4$ & $43/700$ & 1 \\ \hline
   &    1       & 0 &    \\ \cline{2-4}
$M_1$   & $\bar\omega^2r_0^2$ & 0 & \\ \cline{2-4}
   & $\bar\omega^4r_0^4$ & 0 & \\ \hline
   & 1 & 0 &  \\ \cline{2-4}
$E_2$   & $\bar\omega^2r_0^2$ & $12/5$ & 2  \\ \cline{2-4}
   & $\bar\omega^4r_0^4$ & $-16/7$ & 2 \\ \hline
   & 1 & 0 &  \\ \cline{2-4}
$M_2$   & $\bar\omega^2r_0^2$ & 0 &   \\ \cline{2-4}
   & $\bar\omega^4r_0^4$ & $1/60$ & 1 \\ \hline
   & 1 & 0 &  \\ \cline{2-4}
$E_3$   & $\bar\omega^2r_0^2$ & 0 &   \\ \cline{2-4}
   & $\bar\omega^4r_0^4$ & $729/140$ & 3 \\ \cline{3-4}
   &   & $1/2100$ & 1 \\ \hline
   & 1 & 0 &  \\ \cline{2-4}
$M_3$   & $\bar\omega^2r_0^2$ & 0 &   \\ \cline{2-4}
   & $\bar\omega^4r_0^4$ & 0 & \\ \hline
\end{tabular}
\caption{The contribution to the radiated power of the first few
multipole moments. First column: multipole moment in question. The capital
letter is $E,M$ according to whether the mode is electric or 
magnetic. The subscript is the corresponding $l$.  Second column: the
order of the contribution. Third column: the contribution at that
order. Fourth column: the value of $m$ wherefrom this contribution comes.
Note
that there is a factor of $(2/3)q^2\bar\omega^4r_0^2$ common to
all modes which is suppressed in
this table, and that the $E_3$ mode has contributions of the same
order from two different
values of $m$. }
\label{table1}
\end{table}

\newpage

\begin{table} 
\begin{tabular}{||c|c||} \hline
Component & Expression \\ \hline \hline
$Y_{\hat r}^{l+1,lm}$ & $-\sqrt{\frac{l+1}{2l+1}}Y^{lm}$ \\ \hline
$Y_{\hat r}^{l,lm}$ & $0$ \\ \hline
$Y_{\hat r}^{l-1,lm}$ & $\sqrt{\frac{l}{2l+1}}Y^{lm}$ \\ \hline \hline   
$Y_{\hat\theta}^{l+1,lm}$ & $\frac{1}{\sqrt{(l+1)(2l+1)}}\frac{\,\partial
Y^{lm}}{\,\partial\theta}$ \\ \hline
$Y_{\hat\theta}^{l,lm}$ &
$-\frac{1}{\sqrt{l(l+1)}}\frac{m}{\sin\theta}Y^{lm}$ \\ \hline
$Y_{\hat\theta}^{l-1,lm}$ & $\frac{1}{\sqrt{l(2l+1)}}\frac{\,\partial
Y^{lm}}{\,\partial\theta}$ \\ \hline\hline
$Y_{\hat\phi}^{l+1,lm}$ &
$\frac{1}{\sqrt{(l+1)(2l+1)}}\frac{im}{\sin\theta}Y^{lm}$ \\ \hline
$Y_{\hat\phi}^{l,lm}$ & $-i\frac{1}{\sqrt{l(l+1)}}
\frac{\,\partial Y^{lm}}{\,\partial\theta}$ \\ \hline
$Y_{\hat\phi}^{l-1,lm}$ &
$\frac{1}{\sqrt{l(2l+1)}}\frac{im}{\sin\theta}Y^{lm}$ \\ \hline
\end{tabular}
\caption{Pure-orbital vector harmonics in a normalized basis. The first
column is the different components of the pure-orbital harmonics, and the
second column is their value in terms of the scalar spherical harmonics.}
\label{table2}
\end{table}


\begin{thebibliography}{99}
\bibitem{lorenz} The Lorenz gauge is consistent with the multipole
decomposition, in the sense that the Lorenz gauge is satisfied separately
by each multipole mode (see Appendix A)
\bibitem{wronskian}The overall sign of the Wronskian
determinant is determined unambiguously by the requirement that Eq.
(\ref{sol}) would be the solution of the inhomogeneous equation
(\ref{eq}). See Appendix B.
\bibitem{rem-ald} Note that the form of the ALD equation depends on the
choice of metric signature. Our metric signature here is $+2$. For a $-2$
metric signature the minus sign turns into a plus sign. (This happens
because the four-velocity normalization is $u_{\alpha}u^{\alpha}=-1$ if
the signature is $+2$, and $u_{\alpha}u^{\alpha}=1$ if the signature is
$-2$.) The  correct relative sign between the two terms in the ALD
equation is most easily found by checking that the four-force is
perpendicular to the four-velocity.
\end{thebibliography}
\end{document}